\documentclass[useAMS,usenatbib]{mn2e}

\usepackage{graphicx}
\usepackage{url}
\usepackage{rotating}
\title[The double Red Giant Branch in M~2:  C, N, Sr and Ba abundances]{The double RGB in M~2: C, N, Sr and Ba abundances\thanks{This paper 
used data obtained with the MODS spectrographs built with
funding from NSF grant AST-9987045 and the NSF Telescope System
Instrumentation Program (TSIP), with additional funds from the Ohio
Board of Regents and the Ohio State University Office of Research.
The LBT is an international collaboration among institutions in the United States,
Italy and Germany. LBT Corporation partners are: The University of Arizona o
n behalf of the Arizona university system; Istituto Nazionale di Astrofisica, Italy;
LBT Beteiligungsgesellschaft, Germany, representing the Max-Planck Society,
the Astrophysical Institute Potsdam, and Heidelberg University; The Ohio State
University, and The Research Corporation, on behalf of The University of Notre Dame,
University of Minnesota, and University of Virginia.
}}
\author[C. Lardo et al.]{C. Lardo$^{1}$\thanks{E-mail:
carmela.lardo2@unibo.it}, E. Pancino$^{1, 2}$, A. Mucciarelli $^{3}$, M. Bellazzini $^{1}$,
M. Rejkuba $^{4,5}$,\newauthor
S. Marinoni$^{2,6}$, G. Cocozza $^{1}$, G. Altavilla $^{1}$ 
and S. Ragaini $^{1}$\\
$^{1}$ INAF-Osservatorio Astronomico di Bologna, Via Ranzani 1, I-40127 Bologna, Italy\\
$^{2}$ ASI Science Data Center, I-00044, Frascati, Italy\\
$^{3}$ Department of Physics and Astronomy, University of Bologna, Viale Berti Pichat 6/2, I-40127 Bologna, Italy\\
$^{4}$ ESO, Karl-Schwarzschild-Strasse 2, 85748 Garching b. M\"unchen, Germany\\
$^{5}$ Excellence Cluster Universe, Boltzmannstr. 2, D-85748, Garching, Germany\\
$^{6}$ INAF-Osservatorio Astronomico di Roma, via Frascati 33, I-00040, Monte Porzio Catone (RM), Italy}
\begin{document}

\date{Accepted 2013 May 14.  Received 2013 May 14; in original form 2013 April 5}

\pagerange{\pageref{firstpage}--\pageref{lastpage}} \pubyear{2013}

\maketitle

\label{firstpage}

\begin{abstract}
The globular cluster M~2 has a photometrically detected double red giant branch (RGB) sequence. 
We investigate here the chemical differences between the two RGBs in order to gain insight in 
the star formation history of this cluster. The low-resolution spectra, covering the blue spectral range, 
were collected with the MODS spectrograph on the LBT, and analyzed via spectrum synthesis technique. 
The high quality of the spectra allows us to measure C, N, Ba, and Sr abundances relative to iron for 
15 RGB stars distributed along the two sequences. We add to the MODS sample C and N measurements for 35 
additional stars belonging to the blue RGB sequence, presented in Lardo et al. (2012). 
We find a clear separation between the two groups of stars in $s$-process elements as well as C and N content. 
Both groups display a C-N anti-correlation and the red RGB stars are on average richer in C and N with 
respect to the blue RGB. Our results reinforce the suggestion that M2 belongs to the family of globular 
clusters with complex star formation history, together with $\omega$~Cen, NGC~1851, and M~22. 
\end{abstract}


\section{Introduction}\label{introduzione}

The discovery of photometric multiple evolutionary sequences in Galactic Globular glusters (GCs)
color-magnitude diagrams (CMDs) and the presence of distinctive anti-correlations 
also among unevolved stars has conclusively 
shown that these systems host at least two generations of stars 
(see \citealp{grattonREV} and references therein for a discussion).
From a chemical perspective, GCs show a large internal variation in 
the abundances of light elements (Li, C, N, O, F, Na, Mg, and Al; i.e.,  \citealp{kraft94}, 
\citealp{gratton04}, \citealp{carrettaGLOBAL}, \citealp{martell09}, \citealp{kayser08},
\citealp{pancino10}, \citealp{grattonREV}).
On the contrary, the abundances of heavier $\alpha$ (Si, Ca, Ti), iron-peak (Fe, Ni, Cu), 
and neutron-capture elements (Ba, La, Eu) only rarely show similar star-to-star variation.
Besides the remarkable exception of $\omega$~Centauri and Terzan~5 (see \citealp{j10}, \citealp{ferraro09},
and references therein), variations in the heavy element content have been detected only for
few clusters (e.g., \citealp{sneden97,marino11,yong08,villanova10,carretta1851} and references therein).
% \citet{marino09,marino11,marino12} showed that M~22 hosts two metallicity star groups 
% separated by a difference in the iron content of $\simeq$ 0.15 dex, each with its own Na-O and C-N anti-correlation.
% These two metallicity groups are characterized primarily by different relative contents of $s$-process 
% elements. Similarly to M~22, also NGC~1851 shows $s$-process element variations correlated with 
% Al and Na abundances \citep{yong08,villanova10,carretta1851}.
% Their color-magnitude diagrams (CMDs) are also complex, with split at the level of red giant branch (RGB) 
% and subgiant branch (SGB) \citep{marino09,marino11,marino12,milone08,han09,villanova10,carretta1851,lardo12,gratton1851}.

M~2 (=NGC~7089) is an intermediate metallicity ([Fe/H] =--1.62; \citealp{armandroff88})
GC located in the Galactic halo. Its integrated $V$ magnitude ($M_{V}$= --9.02, \citealp{mackey05}) 
puts it among the twelve most luminous clusters
in our Galaxy. As other massive GCs, M~2 shows a peculiar horizontal branch morphology (HB) with a long blue tail 
(see for example Figure~3 of \citealp{dalessandro09}), possibly indicative of significant helium 
variations\footnote{Star-to-star helium abundance variations are expected as the N-Na anomalies are synthetized during 
hydrogen burning.} (see \citealp{dantona02}, \citealp{dantona05}, \citealp{dalessandro13}, and references therein) and a 
split SGB \citep{piotto12}. 

So far only a few spectroscopic studies have been dedicated to M~2 stars.
\citet{armandroff88} reported a metallicity [Fe/H] =--1.62 from integrated-light spectroscopy in the region of the Ca II
infrared triplet. \citet{smith90} found M~2 to have a bimodal CN distribution, 
with the majority of red giants found to be CN-strong stars (but see also earlier results by  
\citealp{mcclure81} and \citealp{canterna82}). M~2 also contains two CH stars \citep{zinn81,smith90}, 
that are relatively rare within GCs (e.g., \citealp{harding62}, \citealp{bond75}, \citealp{mcclure77}, \citealp{hesser82}, 
\citealp{smith82}, \citealp{cote97}, \citealp{sharina12}).
\citet{smolinski11} found that the signs of nitrogen enrichment are present well before the 
point of first dredge-up.

In \citet{lardo11} we were able to detect a spread in light-element abundances of M~2 RGB stars 
using Sloan $u, g, r$ photometry. 
In order to measure the exact content of the C and N of RGB stars, we obtained a number of spectra of giants with the
DOLORES spectrograph\footnote{DOLORES (Device Optimized for the LOw RESolution) is a low resolution spectrograph and camera 
permanently installed at the Nasmyth B focus of the Telescopio Nazionale Galileo (TNG), located in La Palma, 
Canary Islands, Spain.} (\citealp{lardoM2}, hereafter L12). 
We derived C and N abundances from low-resolution blue spectra of 35 giants with magnitude 17.5 $\leq V \leq$ 14.5 mag,
and found that the carbon and nitrogen abundances are anticorrelated, with a hint of bimodality in the C content 
for stars with luminosities below the RGB bump (V$\simeq$15.7 mag), while the range of variations in N abundances is very 
large and spans almost $\sim$ 2 dex.
More interestingly, using $U$ and $V$ images taken during the same observing run, we discovered
an anomalous sequence in the CMD of M~2. 
This feature appears as a narrow, poorly populated RGB, which extends down to the sub giant branch region.
In L12 we speculated that this red RGB could be the extension 
of the faint component of the split SGB recently discovered by \citet{piotto12}.
Unfortunately, no spectra were obtained at that time for stars in
this previously unknown substructure. 

In this paper we try to fill this gap and present a chemical abundance analysis of 
a sample of 15 stars distributed along the double RGB of M~2.
For clarity, in the following we call B-RGB the bluer sequence containing the 
bulk of the RGB population of the cluster, and R-RGB the sparse redder sequence 
identified for the first time in L12.

This article is structured as follows:
we describe the sample in Section~\ref{OSSERVAZIONI};
we outline our measurements of the CN and CH indices and their interpretation in Section~\ref{IND};
we derive C, N, Ba and Sr abundances from spectral synthesis in Section~\ref{ABBONDANZE} and discuss the result in 
Section~\ref{RISULTATI}. 
Finally we present a summary of our results and draw conclusions in Section~\ref{CONCLUSIONI}.

\section{Observations and data reduction}\label{OSSERVAZIONI}
A careful target selection was made for this study.
In order to compare in a homogeneous manner abundances of B-RGB stars with those of stars located on the R-RGB, 
we obtained spectra on both RGBs.

These objects have  magnitude 14 $\leq V \leq$ 17~mag and are present both below and above the bump in 
the RGB luminosity function (see L12 and Sect.~\ref{EV_EFF} for a discussion on the changes on the C and N abundances 
along the RGB evolution).
In particular, we selected M~2 spectroscopic targets from $UV$ photometry presented in L12.
The initial sample of candidate stars consisted of those located more than 
1\arcmin\ and less than 4.5\arcmin\ from the center of M~2 to facilitate sky subtraction and reduce 
fore/background contamination, respectively.
Spectroscopic targets were hence chosen as the most isolated stars\footnote{Only stars of magnitude $V_{0}$ lacking 
neighbors brighter than $V_{0}$+2.0 within 2\arcsec~of their center were kept in the final target list.} 
to avoid the contamination of the spectra from other sources.

\begin{figure*}
\includegraphics[width=17.3cm]{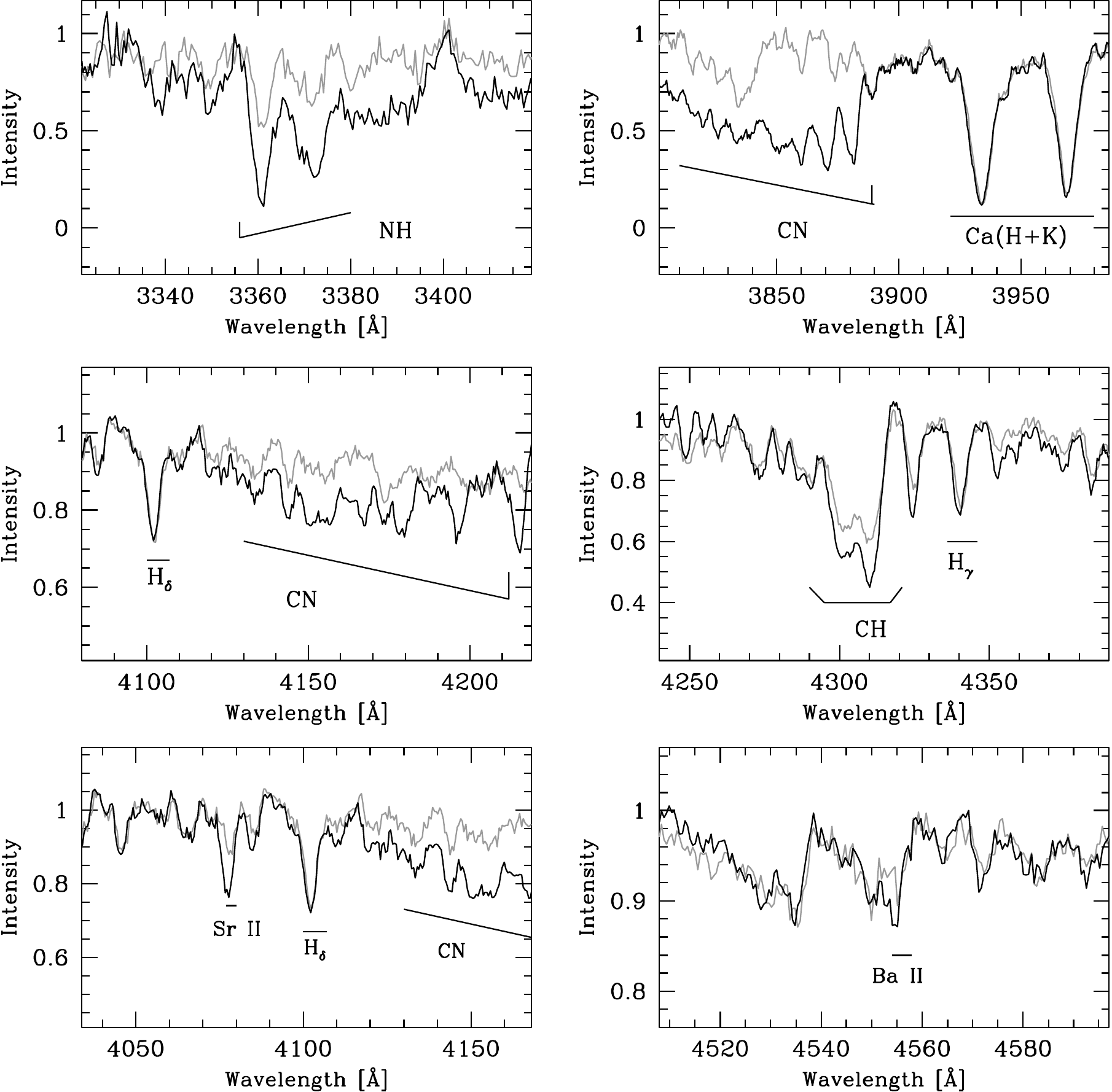}
\caption{{From left to right, top to bottom:} comparison of the spectra of NH (at $\sim$3360~\AA),
CN (at $\sim$3883~\AA~and $\sim$4200~\AA) and CH (at ~4300~\AA) bands, Sr II lines (4077~\AA) and Ba II (4554~\AA) 
in two blue and red RGB stars with similar atmospheric parameters.
The spectrum in black is that of the R-RGB star 05569 (with $T_{eff}$= 5139 K and $\log g$=2.5), 
and the one in gray is that of the B-RGB star 04837 (with $T_{eff}$= 5172 K and $\log g$=2.5).}
        \label{SPETTRO}
   \end{figure*}

The observations were secured with MODS, the low- to medium-resolution Multi-Object 
CCD Spectrograph  operating at the twin 8.4-meter diameter mirror Large Binocular Telescope (LBT)
on Mt. Graham in southeastern Arizona\footnote{{\tt http://www.astronomy.ohio-state.edu/MODS/}}.
MODS allows to allocate slitlets over a 6\arcmin$\times$ 6\arcmin~field of view (FoV).
We defined one slit mask using the stand-alone version of the Multi-Slit Mask Design Software,
provided by the telescope staff\footnote{{\tt http://www.astronomy.ohio-state.edu/~martini/mms/}}. 
Spectra for RGB stars were obtained with the 400 line mm$^{-1}$ reflection grating\footnote{With a 
dispersion of 0.120~\arcsec/pix.}, 
covering a wavelength range of 3200-5800~\AA~and their resolution was R $\simeq$ 800, 950, and 1030 at 
3360, 4000, and 4300~\AA, respectively. 
The slit width on the masks was fixed to
1.0\arcsec, and the slit length was chosen to be at least 12~\arcsec~to allow
for local sky subtraction.
In order to reach high $S/N$ ratio, the mask configuration was observed three times with exposure
durations of 1200 s each, leading to a total exposure time of
1.0 hr and a typical $S/N$ ratio of $\simeq$ 50-60 at 4000~\AA.
Spectra extraction and wavelength calibration were performed by the Italian LBT Spectroscopic 
Reduction Center\footnote{{\tt http://lbt-spectro.iasf-milano.inaf.it}} with a dedicated pipeline.
An example of the spectra quality can be found in Figure~\ref{SPETTRO}, where some of the regions of interest for 
the present analysis are shown.
   
\subsection{Membership}
\begin{figure}
\includegraphics[width=\columnwidth]{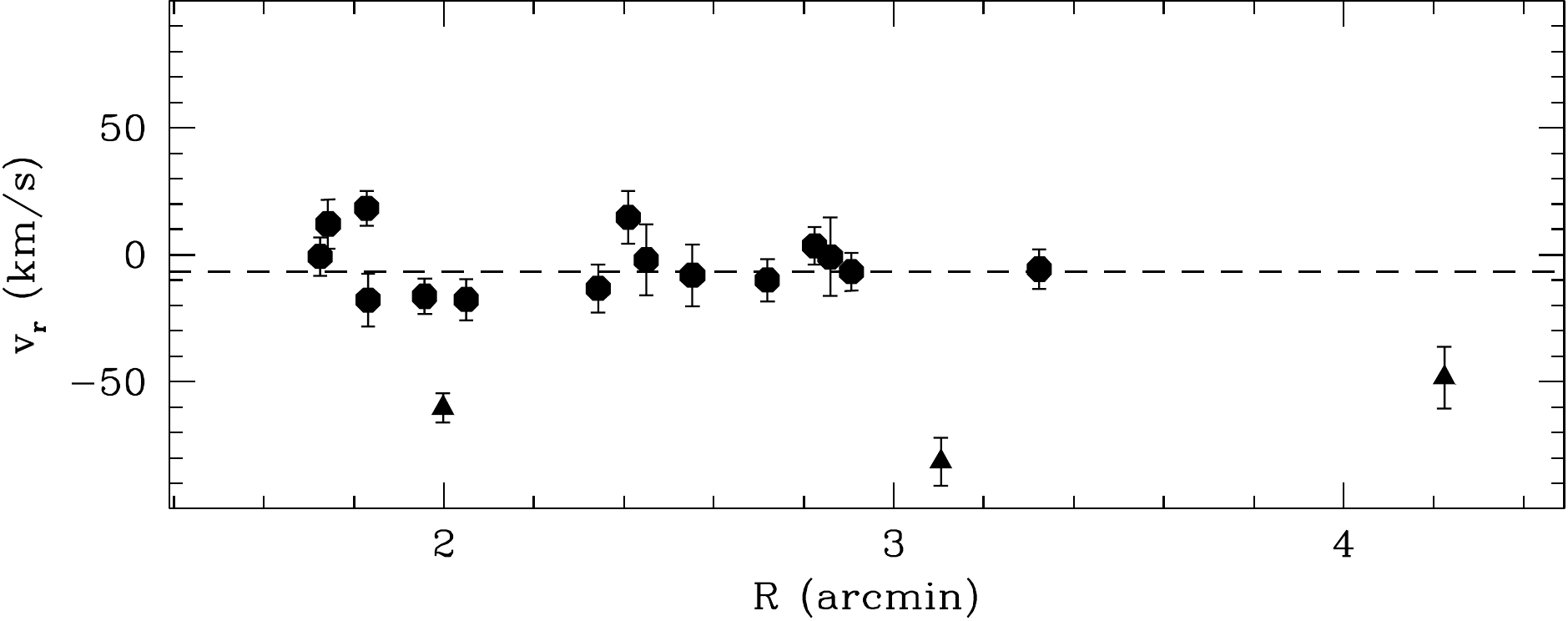}
\caption{Radial velocity plotted against distance from the cluster center for the stars in our sample.
The dashed line is the cluster median velocity of --6.7 km/s. Triangles refer to stars which are not members 
of M~2, according to their radial velocity.}
        \label{VELOCITY}
   \end{figure}

To derive the radial velocity of candidate RGB stars, we first performed a cross-correlation of 
the object spectra with the highest $S/N$ star on each MOS mask as a template with the IRAF routine {\em fxcor}, as 
in L12. The radial velocity of the template star was computed using the laboratory positions of several spectral features
(e.g., H$_{\alpha}$, H$_{\beta}$, H$_{\gamma}$, H$_{\delta}$, and CaII (H+K)).
The measured velocities are plotted as a function of radial distance from the cluster center in Figure~\ref{VELOCITY} and 
listed in Table~\ref{tab_indici}\footnote{For completeness, we report in this Table also the stars judged cluster non-members 
based on their radial velocity.}. The median velocity of --6.7 $\pm$ 8.5 km/s for the entire sample is shown as a dashed line. 
This value, given the low resolution of our spectra, agrees quite well with the value 
tabulated (--5.0 km/s) in the \citealp{harris96} (2010 edition) catalog.
Then, we rejected individual stars with values deviating by more than 3$\sigma$ from this average velocity, 
deeming them to be probable field stars. Three stars (triangles in Figure~\ref{VELOCITY}) were rejected using 
this criterium.
To obtain additional membership information 
we employed the strength of the CaII H and K and H$_{\beta}$ lines, as a further discriminant 
between cluster and field star (see also L12 and Section~\ref{IND}). 
%%%%%%%%%%%%%%%%%%%%%%%%%%%%%%%%%%%%% REFEREE COMMENTS%%%%%%%%%%%%%%%%%%%%%%%%%%%%%%%%%%%%%%%%%%%%%%%%%%

We noted that one more star (located at the blue edge of the RGB sequence)
had discrepant CaII (H+K) and  H$_{\beta}$ with respect to the bulk 
of M~2 stars and rejected it from following analysis.

Yet, since the radial velocity of M~2 stars is near to zero, field stars could easily have such velocities.
As extensively discussed in L12 (see Section~6), we expect a very small degree of contamination given the relatively high 
galactic latitude of the cluster (i.e.; l, b =53$\degr$,-36$\degr$) and the small 
FoV considered to select target stars (see Section~\ref{OSSERVAZIONI}).
In L12 we used TRILEGAL \citep{trilegal} and determined that the fraction of Galactic field stars 
with color 0.4 $\leq$ $U-V$ $\leq$ 2.0, magnitude 18.5 $\leq$ $V$ $\leq$ 14.5 mag, and distance from the cluster
center  1\arcmin $\leq$ R $\leq$ 4\arcmin~is lower than 1\%.
For this work we used both TRILEGAL and the  Besan\c{c}on  model \citep{robin03}\footnote{\url{http://model.obs-besancon.fr/}}.
to obtain a conservative estimate of the degree of contamination affecting our 
spectroscopic sample\footnote{For this test we condidered only stars with 0.7$\leq(U-V)\leq$2.7 and 17.0$\leq V \leq$14.0 mag.}.
We found only 2 (TRILEGAL) or 4 (Besan\c{c}on)\footnote{We note that only 3 out 4 stars have 
velocities comparable to M~2 heliocentric radial velocity.} stars field stars in with $U-V$ color and $V$ magnitude comparable 
with those analyzed in this paper.
All these stars are relatively metal rich dwarfs ($\log$ g $\geq$ 4 dex).
According to TRILEGAL, their metallicity ranges from [Fe/H]= --0.61 dex up to [Fe/H]= -0.13 dex, 
while the Besan\c{c}on model gives a range in metallicity between [Fe/H]= --0.92 and [Fe/H]= 0.13 dex.
Thus our moderate-resolution spectra are in themselves capable of confiming membership.
The spectra of stars rejected on the basis of their radial velocity unambiguously indicates that two of 
these stars display absorption features very different from those shown by the cluster members and 
much stronger than expected for the cluster low metallicty (i.e., [Fe/H]= --1.62 dex). 
One star appears to have metallicity comparable to those of cluster members, although the measured 
radial velocity seems to exlude this.
Consequently, indices and C and N abundances were not measured for it.
To conclude, we note that the stellar spectra shown in Figure~\ref{SPETTRO} are essentially identical everywhere -- and this is
in particular true for the Ca(H+K) lines -- but 
in the NH, CN, and CH absorption regions. This strongly supports the fact that red-RGB stars are indeed M~2 members.

 \begin{figure*}
 \begin{center}
\includegraphics[width=9.2cm, angle=-90]{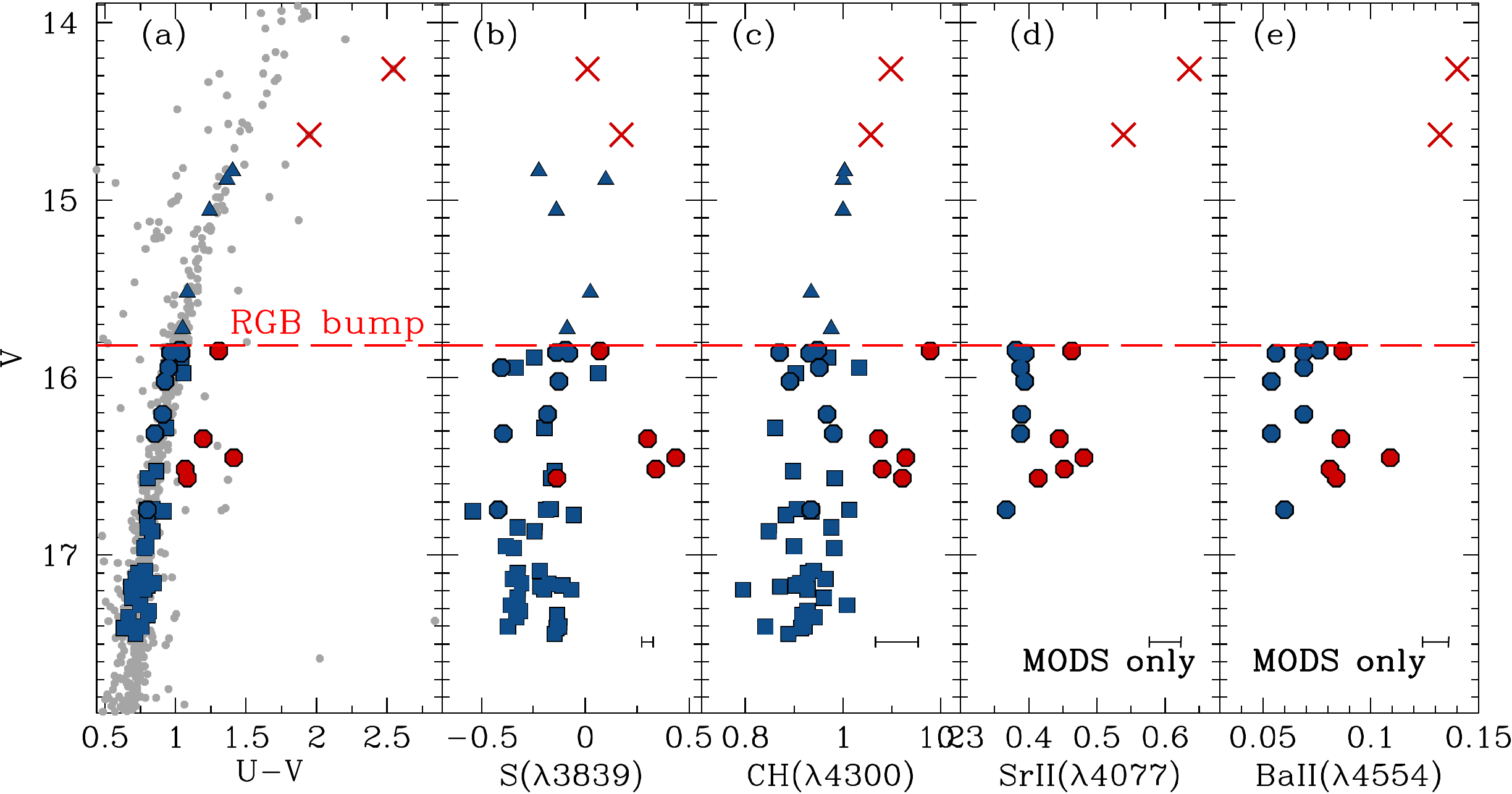}
\vspace{0.5cm}
\caption{In the {\em (a)} panel we present the $V, U - V$ CMD for M~2 from DOLORES images (L12). 
Squares are the stars presented in L12, while circles refer to the stars analyzed in this work.
Stars belonging to the B- and R-RGB are shown in blue and red color, respectively.
The location of the RGB bump is indicated by the horizontal dashed line.
Stars with luminosities brighter that the RGB bump are plotted as crosses (MODS sample) and triangles (DOLORES spectra).
The same symbols and colors are used consistently throughout the whole paper.
The {\em (b)}, {\em (c)}, {\em (d)}, and {\em (e)} panels  present the S($\lambda$3839), CH($\lambda$4300),
SrII($\lambda$4077), and BaII($\lambda$4554) indices plotted against stellar 
$V$ magnitude (with typical measurement error in the bottom right corner), respectively.}

\label{CN}
\end{center}
\end{figure*}

 \section{Spectral indices}\label{IND}
As a preliminary step, we measured a set of indices quantifying the strength of the $UV$ CN band and the G band 
of CH, the CaII H and K, and H$_{\beta}$ lines.
We adopted the same indices as defined in \citet{pancino10} and L12.
The uncertainties related to the index measure have been obtained as in L12 with the expression 
derived by \citet{Vollmann06}, assuming pure photon
noise statistics in the flux measurements.

The measured indices are plotted in Figure~\ref{CN} versus {\it V} magnitude, which shows also the L12 stars 
for comparison. The two stars above the RGB bump are significantly affected by internal mixing and are discussed 
separately in Section~\ref{EV_EFF}.
The indices, together with additional information on target stars, are listed in Table~\ref{tab_indici}.

A visual inspection of the Figure~\ref{CN} {\em (b)} and {\em (c)} panels reveals that the stars in the 
R-RGB sequence have both higher S($\lambda$3839) and CH($\lambda$4300) than stars in the B-RGB. 
The difference in S($\lambda$3839) between CN-strong and CN-weak stars below the bump is 
$\Delta$S($\lambda$3839) $\simeq$ 0.2 mag, while the measurement error is only
of the order of $\simeq$ 0.03 mag.
The Figure~\ref{CN} panel {\em (c)} shows a plot of the CH($\lambda$4300) index vs. the {\it V} magnitude and 
illustrates the relation between the CN and CH band strengths for all giants.
% it shows a plot of the CH($\lambda$4300) index vs. the {\it V} magnitude with the CN-strong and CN-weak stars are also plotted.
In this case the index value spread among the measured stars is very small and 
in any case within the uncertainties. 
% There is a tendency for both groups (i.e., blue and red RGB stars) of CN-strong stars to be also CH-weak, 
% even if exceptions exist.

Because the visual appearance of the MODS spectra suggests enhanced lines of both SrII (4077~\AA)  
and BaII (4554~\AA) -- see also Figure~\ref{SPETTRO}, we measured indices also for these 
elements\footnote{The higher resolution of MODS data allows for this, 
while we cannot provide meaningful index measurements for 
the DOLORES sample.}.
The definition of the band passes is the same as in \citet{stanford07}.
The corresponding indices with median error measurements are shown in the panels {\em (d)} and {\em (e)} of Figure~\ref{CN}. 
Studying the two plots, one can see that R-RGB stars have, in general, considerably larger index values for both SrII(4077~\AA) 
and BaII(4554~\AA)  with respect to the B-RGB stars, indicating that stars belonging to the R-RGB are 
enhanced most probably in $s$-process element content relatively to the B-RGB stars (see also Figure~\ref{SPETTRO}).
We refer the reader to Section~\ref{ELEMENTI_S} for further discussion on this.

   \begin{table*}
   \setlength{\tabcolsep}{0.15cm}
 \begin{minipage}{180mm}  
\caption{Sample stars' properties and index measures.}
\label{tab_indici}
\begin{tabular}{@{}cccccrrrrrrrrrr}
\hline
ID$^{a}$   & RGB$^{b}$&     RA	  &    DEC    &	  V       &  V$_{r}$ & eV$_{r}$ &  CN   &  eCN  &CH & eCH &   SrII & eSrII & BaII & eBaII\\
           &          &   (J200) &   (J200)  &   (mag)   & (km/s)   & (km/s)   &  (mag)& (mag) &(mag) & (mag) & (mag) & (mag) &  (mag) & (mag) \\ 
\hline

05569 & R & 21:33:21.25 & -00:50:57.23 & 16.516 &  -8.1  & 12.2 &  0.341 & 0.029 & 1.080 & 0.039 & 0.452 & 0.023 & 0.081 & 0.006 \\ 
06893 & R & 21:33:24.23 & -00:50:33.21 & 16.566 &   -0.7 &  7.5 & -0.137 & 0.022 & 1.121 & 0.044 & 0.414 & 0.023 & 0.084 & 0.006 \\ 
08017 & R & 21:33:20.09 & -00:50:13.29 & 15.850 &   -2.1 & 14.0 &  0.071 & 0.019 & 1.178 & 0.044 & 0.463 & 0.023 & 0.087 & 0.006 \\ 
08984 & R & 21:33:18.22 & -00:49:56.10 & 14.633 &    3.5 &  7.5 &  0.175 & 0.029 & 1.057 & 0.044 & 0.539 & 0.024 & 0.132 & 0.006 \\ 
10130 & R & 21:33:19.70 & -00:49:36.08 & 16.452 &   14.7 & 10.5 &  0.436 & 0.035 & 1.128 & 0.045 & 0.481 & 0.024 & 0.109 & 0.006 \\ 
14911 & R & 21:33:23.11 & -00:48:11.07 & 14.261 &  -16.4 &  7.0 &  0.011 & 0.033 & 1.098 & 0.065 & 0.636 & 0.025 & 0.140 & 0.006 \\ 
16518 & R & 21:33:32.53 & -00:47:44.36 & 16.345 &  -17.8 & 10.4 &  0.301 & 0.063 & 1.072 & 0.080 & 0.445 & 0.023 & 0.086 & 0.006 \\ 
03285 & B & 21:33:22.72 & -00:51:43.29 & 16.745 &   -0.8 & 15.4 & -0.420 & 0.020 & 0.934 & 0.033 & 0.367 & 0.023 & 0.060 & 0.006 \\ 
04837 & B & 21:33:25.46 & -00:51:11.67 & 16.316 &  -17.7 &  8.1 & -0.395 & 0.016 & 0.980 & 0.025 & 0.388 & 0.023 & 0.054 & 0.006 \\ 
11073 & B & 21:33:19.93 & -00:49:20.21 & 16.021 &  -13.3 &  9.4 & -0.127 & 0.071 & 0.891 & 0.117 & 0.394 & 0.023 & 0.054 & 0.006 \\ 
14279 & B & 21:33:23.09 & -00:48:25.03 & 16.206 &   18.3 &  6.8 & -0.182 & 0.019 & 0.967 & 0.027 & 0.390 & 0.023 & 0.069 & 0.006 \\ 
15906 & B & 21:33:20.10 & -00:47:55.92 & 15.863 &  -10.0 &  8.3 & -0.079 & 0.018 & 0.931 & 0.030 & 0.395 & 0.023 & 0.056 & 0.006 \\ 
18369 & B & 21:33:22.04 & -00:47:06.84 & 15.944 &   -6.7 &  7.5 & -0.404 & 0.018 & 0.951 & 0.035 & 0.388 & 0.023 & 0.069 & 0.006 \\ 
20166 & B & 21:33:23.46 & -00:46:23.85 & 15.845 &   -5.7 &  7.7 & -0.095 & 0.017 & 0.948 & 0.024 & 0.381 & 0.023 & 0.076 & 0.006 \\ 
11876 & B & 21:33:22.42 & -00:49:06.09 & 15.859 &   12.1 &  9.6 & -0.138 & 0.018 & 0.870 & 0.030 & 0.384 & 0.023 & 0.069 & 0.006 \\ 
04087 & NM & 21:33:14.53 & -00:51:26.07 & 14.800 &  -48.5 & 12.2 &$\cdots$ &$\cdots$&$\cdots$&$\cdots$&$\cdots$&$\cdots$&$\cdots$&$\cdots$ \\ 
17531 & NM & 21:33:30.33 & -00:47:24.12 & 15.278 &  -60.3 &  5.8 &$\cdots$ &$\cdots$&$\cdots$&$\cdots$&$\cdots$&$\cdots$&$\cdots$&$\cdots$ \\ 
19000 & NM & 21:33:36.69 & -00:46:53.30 & 16.748 &  -81.6 &  9.4 &$\cdots$ &$\cdots$&$\cdots$&$\cdots$&$\cdots$&$\cdots$&$\cdots$&$\cdots$ \\

\hline
\end{tabular}

\medskip
$^{a}$ ID is star ID number from \citet{lardoM2} photometric catalog.\\
$^{b}$ Stars are defined B or R according to their location on the B- or R-RGB, respectively, in Figure~\ref{CN}.
We flag M~2 non-member stars with NM. 

\end{minipage}
\end{table*}

 \section{Abundance analysis}\label{ABBONDANZE}
 To obtain a quantitative estimate of the detected enhancements in carbon, nitrogen and strontium for R-RGB stars, we derived
 [C/Fe], [N/Fe], [Sr/Fe], and [Ba/Fe] abundance ratios for all MODS stars via spectral synthesis.
 \subsection{Atmospheric parameters}\label{PARAMETRI_ATM}
Stellar parameters were derived from photometry, in the same fashion as described in L12.
Briefly, the effective temperature, $T_{\rm{eff}}$, 
was calculated using the \citet{alonso99}
$T_{\rm{eff}}$-color calibrations for giant stars. 
We used the $(U-V)$ color from 
DOLORES photometry (once calibrated on Stetson standard field), using
 $E(B-V) =0.06$  and [Fe/H] =--1.65 from the \citet{harris96} catalog (2010 edition). 
 In addition, we used --- when available --- $(B-V ), (V-J ), (V-H ),$ and $(V-K )$ colors 
 from \citet{lee99} and the 2MASS photometry\footnote{{\url http://www.ipac.caltech.edu/2mass/overview/access.html}}. 
 The final $T_{\rm{eff}}$ was the mean of the individual $T_{\rm{eff}}$ values from each color, weighted 
 by the uncertainties of each color calibration. 
 The surface gravity was determined using $T_{\rm{eff}}$ , 
 a distance modulus of $(m-M)_{V}$=16.05 \citep{harris96}, 
 bolometric corrections BC(V) from \citet{alonso99} and assuming a mass of 0.8 $ M_{\odot }$ \citep{berg01} . 
 The microturbulent velocity was determined using $v_{t}=-8.6 \times 10^{-4} T_{\rm{eff}} + 5.6$, 
 adopted from the analysis by \citet{pila96} of metal-poor subgiant and giant stars 
 with comparable stellar parameters. 
 Table \ref{tab_abundances} reports the $T_{\rm{eff}}$, $\log g$ and $v_{t}$ values and their 
 uncertainties; these values are used to choose the atmospheric model for the spectral synthesis of each star.
 
  \begin{table*}
 \begin{minipage}{180mm}
\setlength{\tabcolsep}{0.18cm}

\caption{Atmospheric parameters and C, N, Ba, and Sr abundance ratio for the M~2 cluster members.}
\label{tab_abundances}
\begin{tabular}{@{}lccccccrrrrrrrr}
\hline
ID   &  T    &  eT  &  log g &  elog g &   v$_{t}$ & ev$_{t}$ &  [C/Fe] &  e[C/Fe] & [N/Fe]  &  e[N/Fe] &  [Sr/Fe] & e[Sr/Fe] & [Ba/Fe] & e[Ba/Fe] \\ 
     & (K)   &  (K) & (dex)  & (dex)   &   (km/s)  & (km/s)   &   (dex) & (dex)    & (dex)   &   (dex)  &  (dex)   &  (dex)   &  (dex)  & (dex)   \\ 
\hline

05569 & 5136 & 59 & 2.5 & 0.03 & 1.3 & 0.4 & --0.37 & 0.22 &  1.30 & 0.23 &  0.85 & 0.22 & 1.22 & 0.28 \\
06893 & 5115 & 65 & 2.5 & 0.03 & 1.2 & 0.3 & --0.24 & 0.18 &  0.61 & 0.25 &  0.45 & 0.23 & 1.15 & 0.30 \\
08017 & 4868 & 67 & 2.2 & 0.03 & 1.4 & 0.5 & --0.23 & 0.19 &  0.47 & 0.23 &  0.77 & 0.21 & 1.09 & 0.29 \\
08984 & 4455 & 42 & 1.4 & 0.03 & 1.6 & 0.6 & --0.99 & 0.20 &  0.51 & 0.22 &  0.81 & 0.21 & 1.12 & 0.31 \\
10130 & 5019 & 55 & 2.5 & 0.03 & 1.3 & 0.4 & --0.29 & 0.20 &  0.94 & 0.23 &  0.97 & 0.21 & 1.27 & 0.28 \\
14911 & 4170 & 46 & 1.1 & 0.04 & 1.8 & 0.7 & --0.93 & 0.19 &  0.07 & 0.23 &  0.53 & 0.22 & 0.88 & 0.28 \\
16518 & 4833 & 51 & 2.3 & 0.03 & 1.4 & 0.5 & --0.56 & 0.20 &  0.88 & 0.23 &  0.75 & 0.21 & 0.97 & 0.29 \\
03285 & 5094 & 59 & 2.6 & 0.03 & 1.3 & 0.4 & --0.49 & 0.19 & --0.39 & 0.26 &  0.26 & 0.21 & 0.50 & 0.31 \\
04837 & 5172 & 79 & 2.5 & 0.03 & 1.3 & 0.4 & --0.63 & 0.18 & --0.15 & 0.23 & --0.12 & 0.22 & 0.34 & 0.30 \\
11073 & 5095 & 78 & 2.3 & 0.03 & 1.3 & 0.4 & --0.93 & 0.19 &  0.85 & 0.23 &  0.35 & 0.22 & 0.59 & 0.32 \\
14279 & 5153 & 78 & 2.4 & 0.03 & 1.3 & 0.4 & --0.52 & 0.19 &  0.72 & 0.23 &  0.25 & 0.22 & 0.57 & 0.35 \\
15906 & 4962 & 73 & 2.2 & 0.03 & 1.4 & 0.5 & --0.90 & 0.19 &  0.83 & 0.23 & --0.10 & 0.21 & 0.52 & 0.29 \\
18369 & 4955 & 73 & 2.2 & 0.03 & 1.4 & 0.5 & --0.79 & 0.19 & --0.31 & 0.23 &  0.15 & 0.21 & 0.33 & 0.30 \\
20166 & 4910 & 72 & 2.2 & 0.03 & 1.4 & 0.5 & --0.91 & 0.18 &  0.63 & 0.23 & --0.04 & 0.21 & 0.66 & 0.27 \\
11876 & 5180 & 60 & 2.3 & 0.03 & 1.3 & 0.4 & --0.66 & 0.18 &  0.61 & 0.24 &  0.35 & 0.22 & 0.45 & 0.30 \\
\hline
\end{tabular}
\end{minipage}
\end{table*}

\subsection{Abundances derivation and error analysis} 
Synthetic spectra were generated using the local thermodynamic equilibrium (LTE) 
program MOOG \citep{sneden73}.
The atomic and molecular line lists were taken from the latest Kurucz compilation \citep{castelli04} and 
downloaded from F. Castelli's website\footnote{\url{http://wwwuser.oat.ts.astro.it/castelli/linelists.html}}.
Model atmospheres were calculated with the ATLAS9 code 
starting from the grid of models available in F. Castelli's website \citep{castelli03}, 
using the values of $T_{\rm{eff}}$, $\log g$, and $v_{t}$ determined as explained in the previous section and 
listed in Table~\ref{tab_abundances}.
For all the models we adopted $[A/H]=-1.5$, according to the metallicity of the cluster.
The ATLAS9 models employed were computed with the new set of opacity distribution functions
\citep{castelli03} and excluding approximate overshooting in 
calculating the convective flux.

C and N abundances were estimated by spectral synthesis of the CH band at
$\sim$4310~\AA~and the NH and CN bands at ~3360 and 3883~\AA, respectively.

\begin{figure}
\includegraphics[width=\columnwidth]{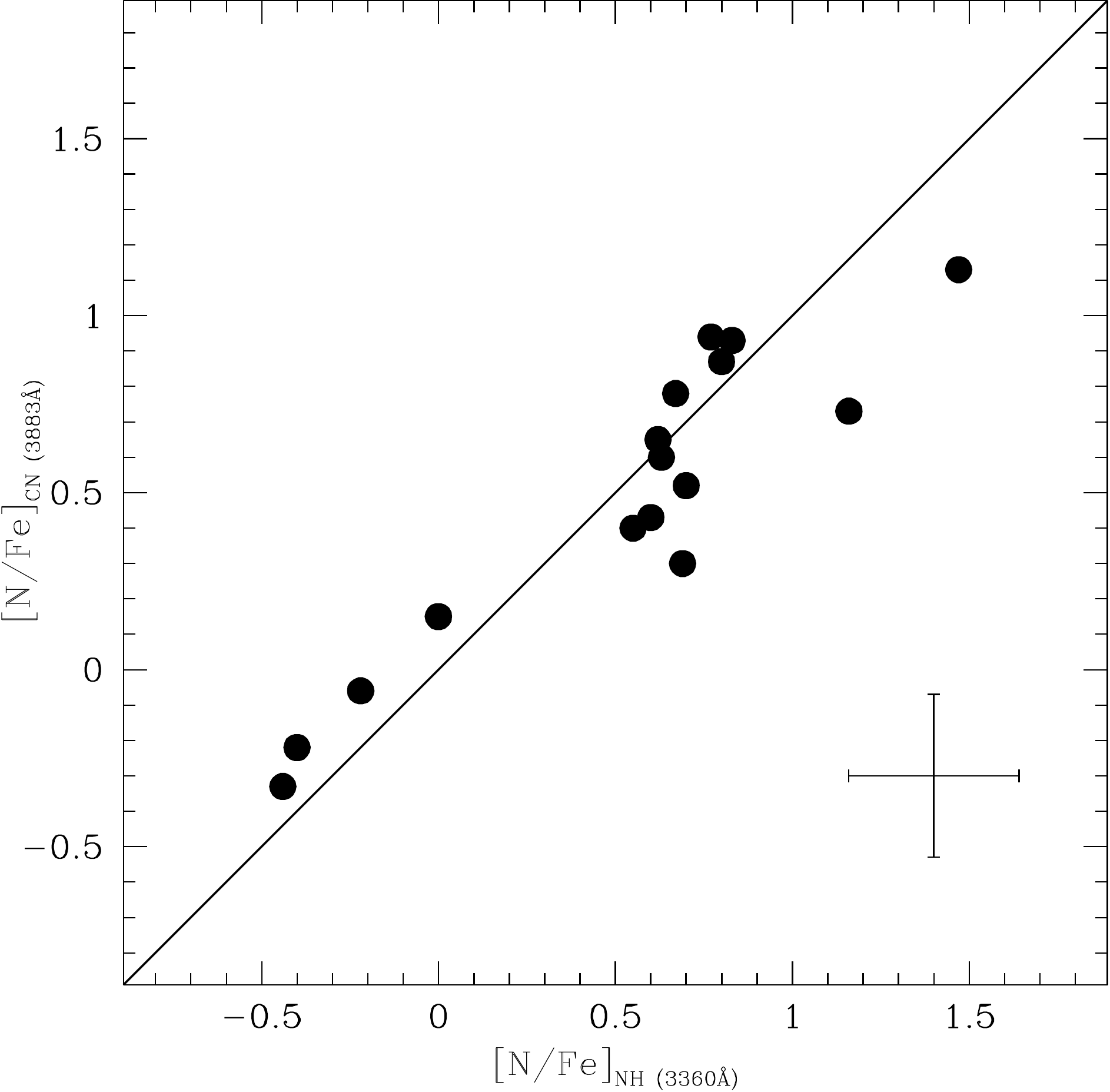}
\caption{Comparison between the nitrogen abundances derived by synthesis of the NH band ($\sim$ 3360~\AA) and those derived from the 
ultraviolet CN band at $\sim$ 3883~\AA. The 1:1 correlation line is also reported.
Typical error bar on the measurements is also shown.}
        \label{CONFRONTO}
   \end{figure}

Figure~\ref{CONFRONTO} compares the N abundances derived from the NH and CN molecular bands.
The correlation is very good and the difference between the two determinations is always within the errors.
Therefore we considered in the following the {\em average} nitrogen abundances obtained from the 
NH and CN molecular bands.
For the CH transitions, the $\log g_{f}$ obtained from the Kurucz database were revised downward by
0.3 dex to better reproduce the solar-flux spectrum by \citet{neckel84}
with the C abundance by \citet{caffau11}, as discussed in \citet{mucciarelli12}.

% To double check the nitrogen abundances results, we derived also N abundance from the CN violet band at $\simeq$3883~\AA.

Finally, the Sr II 4077~\AA~and BaII 4554~\AA~features were analyzed.
% The Sr II 4215~\AA~line was not analyzed owing to contamination 
% from the 4215~\AA~CN features. 
The $gf$-values and solar abundances used for Sr and Ba are the same listed in \citet{stanford07}.
% We investigated the effect of hyperfine splitting (hfs) on the derived abundance from the Ba II 4554 \AA line 
% following \citet{stanford10}. Two line lists were used: one that included hfs and isotopic components for
% Ba ii 4554 Å, and one that did not. 
% A comparison of abundances was made for a series of stellar parameters and 
% [Ba/Fe] spanning the same range as for the RGB sample. 
% Little difference (?[Ba/Fe] < 0.05) was found between the abundances obtained 
% from the spectra with and without the inclusion of isotopes and hfs.

In Figure~\ref{SPETTRO} we show two representative B- and R-RGB stars spectra (gray and black color, respectively) in the 
in NH, CN, CH, Sr II and Ba II spectral regions.

% These stars have essentially the same stellar parameters ($T_{eff} \simeq$5100 $\log$ g =2.5)
% yet all these spectral features differ strongly, the red RGB stars clearly displaying strong NH, CN, and CH 
% absorptions together with enhancement in the Sr and Ba abundances.

An error analysis was performed by varying the temperature, gravity, metallicity, and microturbulence, 
and redetermining the abundances for three representative stars.
Typically for the temperature, we found $\delta$A(C) / $\delta T_{eff} \simeq$ 0.09 - 0.11 dex  
and $\delta$A(N) / $\delta T_{eff} \simeq$ 0.13 - 0.15 dex.
The errors in abundances due to uncertainties on gravity and microturbulent velocity are negligible (on the order of 0.02 dex or less).
We also assumed that all stars had the same oxygen abundance ([O/Fe]=+0.4 dex) regardless of luminosity.
The derived C abundance is thus dependent on the O abundance.
To quantify the sensitivity of the C abundance on the adopted O abundance we varied the oxygen abundances 
and repeated the spectrum synthesis to determine the exact dependence for three representative stars 
(4900 K $\leq T_{eff} \leq$ 5200 K). 
We found that variations in the oxygen abundance 
cause a variation in the derived [C/Fe]  by as much as $\simeq$0.10 - 0.11 dex for a 
0.4 dex change in assumed [O/Fe].
This error was then included in the final error of [C/Fe].

% In addition to the stellar parameters and oxygen abundance errors, a measurement uncertainty exists in 
% the determination of the individual abundances. 
% This was estimated by visual inspection of the quality of the fit for each star for carbon, nitrogen, strontium, and barium, 
% and combined with the general errors above.

All these individual errors were added in quadrature and gave a final error 
$\Delta$[C/Fe] = 0.19 dex, $\Delta$[N/Fe] = 0.23 dex.
The uncertainties in Sr and Ba abundances were determined to be $\Delta$[Sr/Fe] = 0.21 dex and $\Delta$[Ba/Fe] =0.30 dex.

We present the abundances derived as described above and the relative uncertainties in Table~\ref{tab_abundances}. 

 \section{Abundance results}\label{RISULTATI}  

\subsection{C-N anti-correlation}

\begin{figure}
\includegraphics[width=\columnwidth]{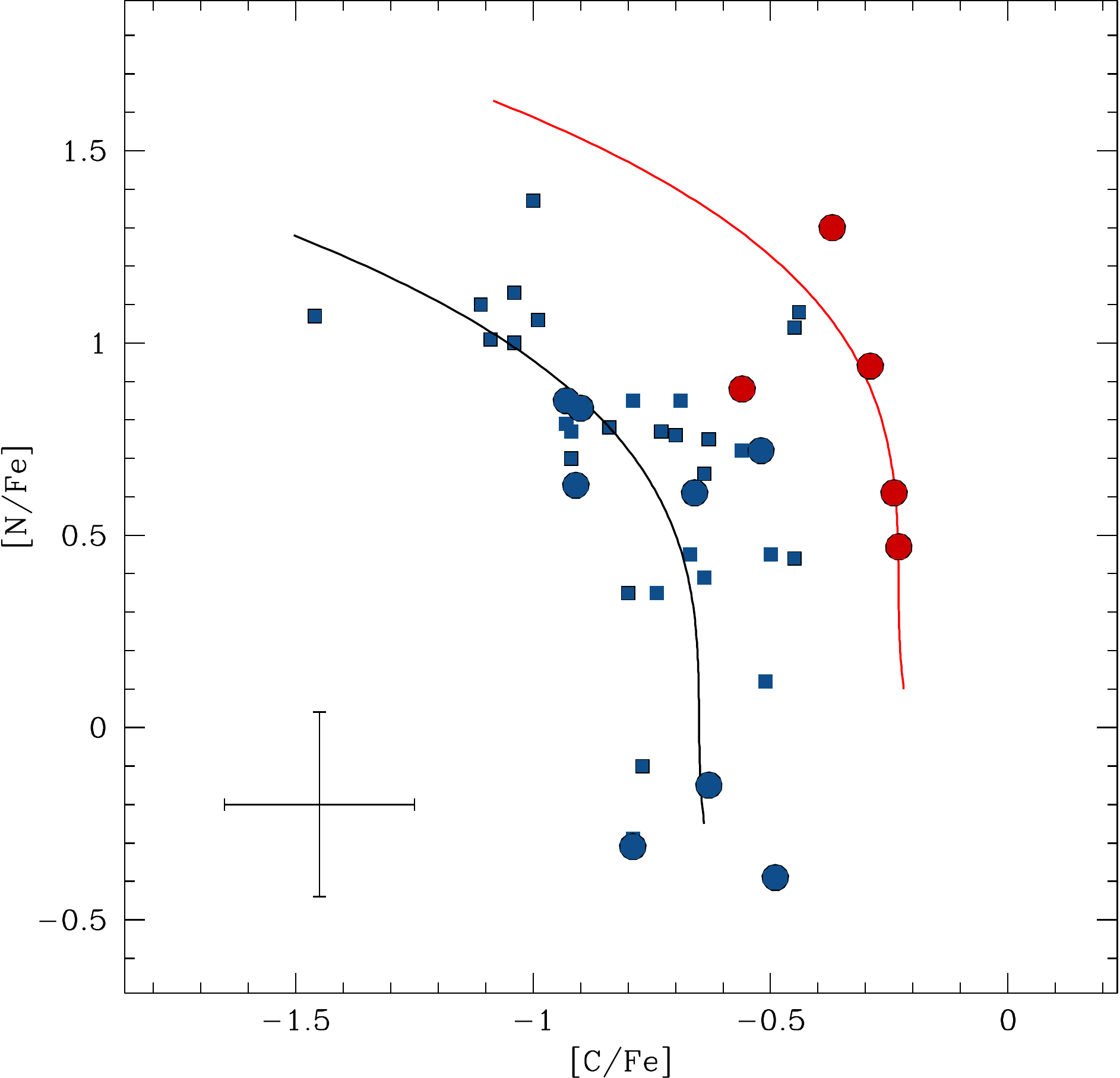}
 
\caption{C and N anti-correlation for all the stars in the DOLORES and MODS sample. For readability purpose we report 
[C/Fe] and [N/Fe] abundance ratios only for stars fainter than the RGB bump.
The black solid line indicate the relationship that prevails in B-RGB stars --with 
equation  [C/Fe]= --0.428 [N/Fe]$^{3}$ +0.034 [N/Fe]$^{2}$ --0.007 [N/Fe] --0.650.
We shifted this fiducial line by $\Delta$[C/Fe]= +0.42 and $\Delta$[N/Fe]= +0.35 (red solid line) to visually fit the 
C-N anti-correlation showed by R-RGB stars. The symbols are the same as in Figure~\ref{CN}.}
 \label{ANTI}
\end{figure}
The derived carbon and nitrogen abundances are shown in Fig.~\ref{ANTI},
where the [C/Fe] values are plotted as a function of [N/Fe]. 
In the same figure we plot also carbon and nitrogen abundances from DOLORES data derived in L12 with 
the [C/Fe]-[N/Fe] relationship that prevails for these stars, to unambiguously identify the locus of 
the bulk of B-RGB stars in the [C/Fe] vs. [N/Fe] space\footnote{We neglect  
stars brighter than the RGB bump because of evolutionary effects on C 
and N surface abundances which will be discussed in Section~\ref{EV_EFF}.}.

Having used the same analysis techniques, the two samples appear comparable (within the errors).
Therefore in the following we discuss the MODS abundances together with DOLORES data presented in L12.

The newly derived C and N abundances for B-RGB stars coincide with the main C-N anti-correlation 
from DOLORES data, while R-RGB stars have, in general, higher carbon 
abundances.
We found that B-RGB stars have on average  [C/Fe] $\simeq$ --0.77 $\pm$ 0.14 dex, that significantly differs from the 
average carbon content of the R-RGB populations ([C/Fe] $\simeq$ --0.29 $\pm$ 0.06 dex).
Again, we noted a fully extended C-N anti-correlation for the B-RGB stars, 
while R-RGB stars tend to be richer in N, on average.
Concerning the nitrogen abundances, we found an average value [N/Fe] $\simeq$ 0.75 $\pm$ 0.20 dex and [N/Fe] $\simeq$ 0.88 $\pm$ 0.16 dex
for the B- and R-RGB population, respectively.
Figure~\ref{ANTI} suggests also that that the R-RGB stars exhibit their own C-N anti-correlation -- although we 
are considering only five stars.
The C-N pattern observed for the RGB stars recalls the Na-O anti-correlation analyzed for RGB stars in 
M~22 \citep{marino09,marino11,marino12}, NGC~1851 \citep{yong08,carretta1851,villanova10} and 
$\omega$~Cen \citep{j10,marinOmega}.
Specifically, the blue and red RGB samples are not completely superimposed on one another 
in the [C/Fe]-[N/Fe] plane; while red RGB stars possibly show a shorter anti-correlation, with, 
on average, a higher nitrogen abundance (see also the behavior of faint and bright SGB stars in NGC~1851 shown in 
Figure~15 by \citealp{lardo12}). 

As discussed in L12, M~2 contains two CH stars (see \citealt{zinn81} and \citealt{smith90}).
These stars show abnormally high CH absorption, together with deep CN bands, compared to other cluster giants.
In our previous work on RGB stars in M~2, we confirmed that both CH stars belong to the additional RGB, 
pointing out the anomalous chemical nature of this redder branch. 
With the present analysis we are able to demonstrate that indeed R-RGB stars are C and N enhanced 
with respect to B-RGB ones.

\subsection{Some comments on the C+N+O sum}

\citet{cassisi08} and \citet{ventura09} argued that the SGB split
of NGC~1851 can be accounted for by a difference in the overall CNO abundance 
between the two SGB groups without assuming a large age difference.
Indeed, \citet{yong09} found evidence for strong CNO variations in a sample of four luminous RGB stars
(but see the result of \citealp{villanova10}).
In \citet{lardo12} we analyzed the C+N sum for SGB stars NGC~1851 and 
measured a difference in log$\epsilon$(C+N) of 0.4 between the two SGBs, 
that implies that the fainter SGB has about 2.5 times the C+N content 
of the brighter one\footnote{However, we did not measure [O/Fe] abundances for our SGB stars and cautioned that 
the separation one sees in C+N content could significantly decrease or disappear when 
considering the C+N+O sum.}.

In M~22 the faint SGB stars are enriched in $s$-process elements and 
additionally in the total C+N+O and metallicity  \citet{marino12}. By accounting for the chemical content of the double SGB, 
these authors found that these two SGB groups do not differ in age by more than $\simeq$ 300 Myr.

\begin{figure}
\includegraphics[width=\columnwidth]{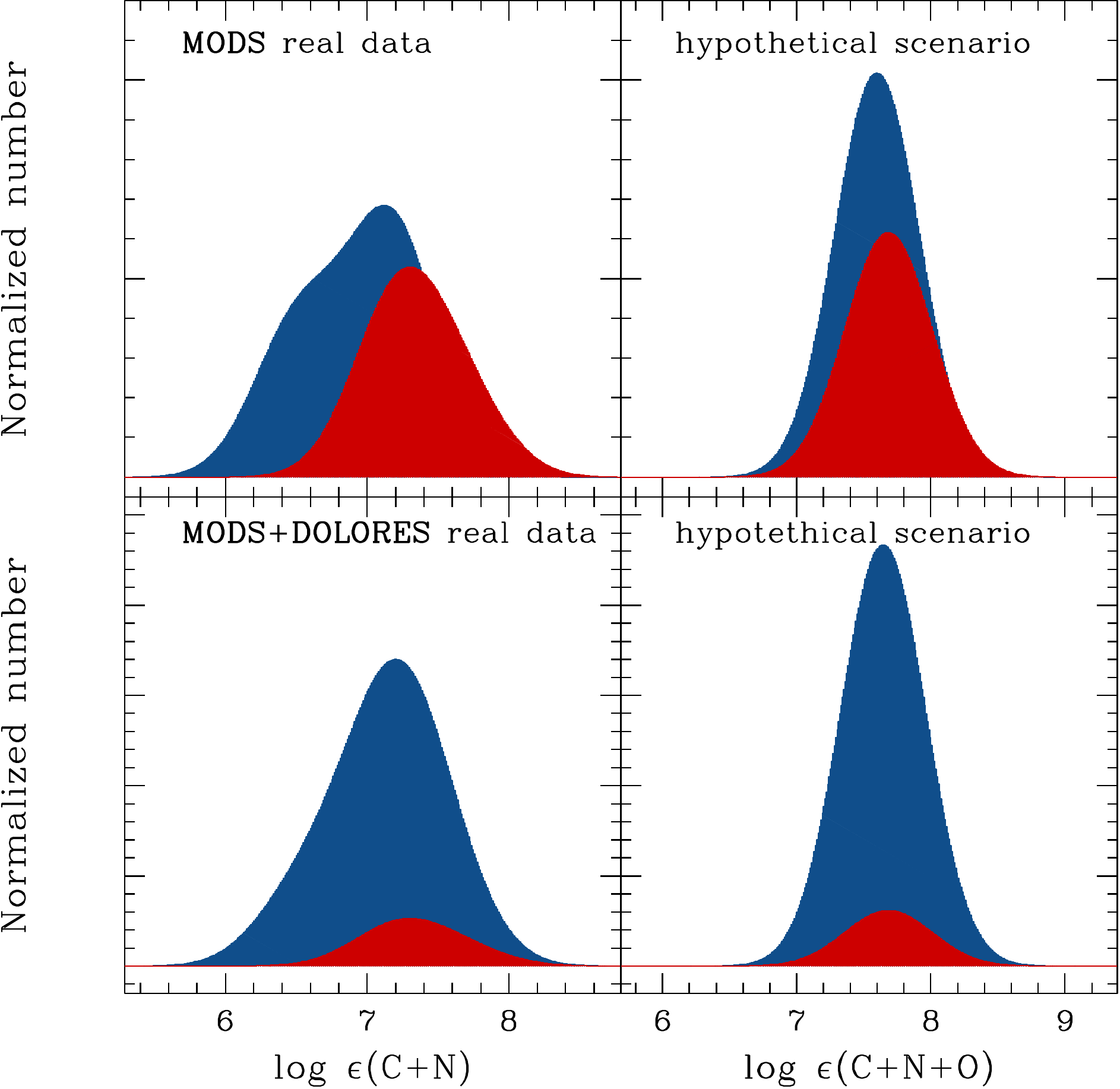}
\caption{{\em Top:} on the left hand panel we reported the generalized histogram of the C+N distribution for MODS stars, flanked by the 
generalized histogram of the C+N+O distribution, obtained by assuming a reference [O/Fe] abundance for the N-rich and N-poor stars 
(see text). {\em Bottom:} the same as in the top panels but for the entire spectroscopic sample (DOLORES + MODS).}
        \label{CNO}
   \end{figure}

To investigate the possibility that the split SGB and the double RGB in M~2 could be due to a different C+N+O content 
between the two subpopulations, we computed the C+N sum for our B- and R-RGB stars. 
In the upper panels of Figure~\ref{CNO}, we plot the generalized histograms of the C+N distribution
for the MODS dataset considering the 
two RGBs separately\footnote{Each data point in these histograms has been replaced by a 
Gaussian of unit area and a standard deviation
$\sigma$=0.30, the typical error associated to the measurements.}.
The average C+N content of the two RGB populations is different: the 
R-RGB has $\log \epsilon$(C + N) $\simeq$ 7.35 $\pm$ 0.25 and the B-RGB has 
$\log \epsilon$(C + N) $\simeq$ 6.93 $\pm$ 0.35. 
When considering also the DOLORES sample, we established that the small difference in the total C+N+O content 
is slightly diminished; in this case the B-RGB has $\log \epsilon$(C + N) $\simeq$ 7.12 $\pm$ 0.33
(top right-hand panel of Figure~\ref{CNO}).

Assuming an N-O anti-correlation, N-poor (first generation) stars in each RGB group
are more O-rich than the N-rich (second generation) stars\footnote{In the following, we arbitrarily
discriminate between first and second generation stars by assuming a threshold in the nitrogen content 
[N/Fe]$\simeq$ 0.6 dex.}. 
Even though we cannot measure oxygen abundances for our RGB stars, we can 
speculate on the C+N+O sum for the two RGB components assuming reasonable [O/Fe] value.
As mentioned in Sect.~\ref{introduzione} no oxygen abundance determination of M~2 RGB stars can be 
found in the literature, therefore we took as reference [O/Fe] values those derived by \citet{marino11} in the case of M~22,
a cluster with similar metallicity ([Fe/H]= --1.70; \citealp{harris96}, 2010 edition).
The latter authors presented [C/Fe], [N/Fe], [Na/Fe] and [O/Fe] abundances for a number of 
RGB stars in M~22 located on both the red and blue RGB in M~22.
Firstly, we note from their Figure~16 that also in that case each RGB displays its own
C-N anticorrealtion, with red RGB stars also having an excess of nitrogen.
Second, again from their Figure~16, we derived a reference [O/Fe] abundance for N-poor and N-rich stars.
We caution readers that assigning a reference [O/Fe] content to each group could be na\"{i}ve 
at this stage\footnote{Moreover, it is highly probable that oxygen abundance differences exist also 
between the R- and B-RGB stars.}. If we assume for the N-poor and N-rich group [O/Fe] = 0.4 dex and [O/Fe] = 0.0 dex, respectively, 
the separation one sees in C+N content is largely diminished when considering C+N+O
(right panels of Figure~\ref{CNO}).
We found that the R-RGB have $\log \epsilon$(C + N + O) $\simeq$ 7.69 $\pm$ 0.13
and the B-RGB $\log \epsilon$(C + N + O) $\simeq$ 7.59 $\pm$ 0.10 and  7.65 $\pm$ 0.10 -- when considering 
only MODS and the entire sample, respectively.
This exercise is not conclusive, of course, and new data on oxygen are badly needed, but it shows that if the analogy 
between M~2 and M~22 holds, the C+N+O sum in M~2 could change by a small amount, and age could play a r\^{o}le.

    \begin{figure}
\includegraphics[width=\columnwidth]{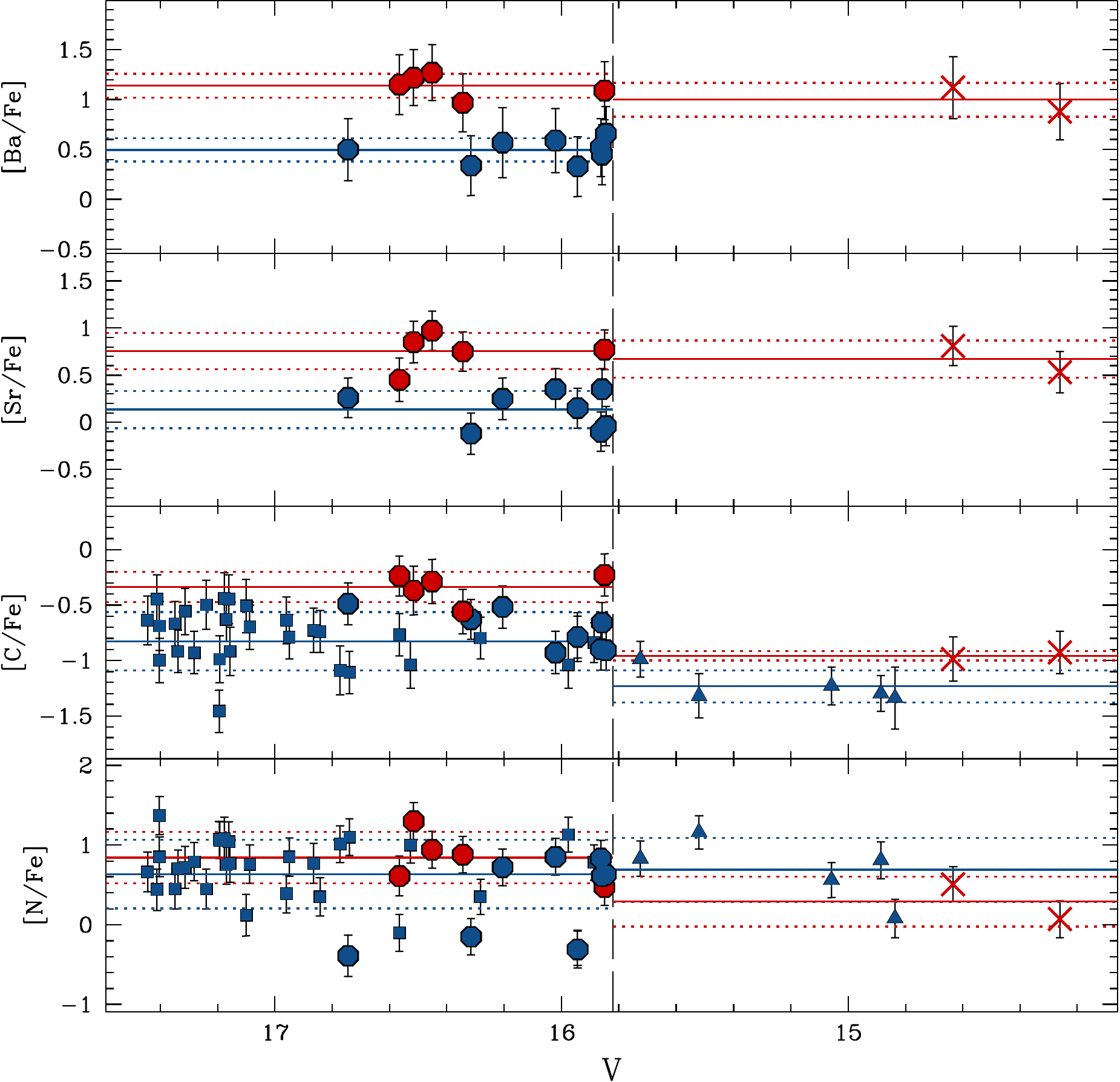}
\caption{Run of the derived abundances with the temperature. The red (blue) horizontal line marks the position of the average 
abundance with the $\sigma$ interval (dotted lines) for the R- (B-) RGB groups, respectively.
The symbols are the same as in Figure~\ref{CN}.}
        \label{TEMPERATURA}
   \end{figure}  
 \subsection{Evolutionary effects}\label{EV_EFF}
In Figure~\ref{TEMPERATURA} we plot the derived abundances against the temperature of stars to evaluate systematic effects with
luminosity (and temperature) and note that neither of these effects are apparent.
Once again we discriminated between stars fainter and brighter than the RGB bump, as significant CNO surface abundance changes
are not expected to occur in stars fainter than the RGB bump \citep{iben68}.
While the Sr and Ba abundances of the R-RGB group are left unchanged as the stars evolve toward brighter luminosities,
Figure~\ref{TEMPERATURA} illustrates the notable depletion in the carbon
abundances with luminosity \citep[][and references therein]{smith03,gratton00}.
%The surface carbon abundance depletion along the RGB of M~2 can be simply interpreted within a deep-mixing 
%framework (i.e.: \citealt{charbonnel95,charbonnel98,angelou12}, and references therein).
This implicates that some form of deep mixing (i. e., meridional circulation currents, turbolent diffusion or some
similar processes; see \citealt{charbonnel95,charbonnel98,angelou12}, and references therein)
must circulate material from the base of the convective envelope down 
into the CN(O)-burning region near the hydrogen burning shell.
In L12, we noted that the decline in the carbon abundance approximately occurred at magnitude $V\simeq15.7$ mag, 
which is essentially the location of the RGB bump in this cluster 
(see L12 for a discussion). Our new data fully confirm this finding.
Restricting our sample to those giants fainter than the RGB bump, we derive an average C abundance of  
[C/Fe]=--0.83$\pm$0.26 dex, while
stars more luminous than the RGB bump have average carbon abundance of [C/Fe]=--1.23$\pm$0.17 dex\footnote{For comparison, 
in metal-poor field giants \citep{gratton00} a drop of the surface $^{12}$C abundance by about a factor 2.5, 
it is seen after this second mixing episode.}.

Conversely, we see no significant trend of the N
abundance with luminosity: the average nitrogen abundance we measure for stars fainter than the RGB bump 
([N/Fe]= 0.63 $\pm$ 0.43 dex) agrees quite well with that obtained
for the more luminous stars after the RGB bump, [N/Fe]= 0.56 $\pm$ 0.38 dex\footnote{We caution, however, that the extent of 
the carbon (nitrogen) depletion (enhancement) depends on the value of [O/Fe] used in the analysis.}.
\citet{gratton00} show for field giants an abrupt increase in N abundance of about a factor of 4 at $\simeq V_{BUMP}$.
Here, as in L12, we could not detect such trend. 
On the other hand, a few studies do not report a significant N enhancement for stars brighter than the RGB bump.
\citet{tref83} studied a sample of 33 bright giants in M~15, a very metal poor cluster ([Fe/H]=-2.37, \citealp{harris96}).
They found that the mean carbon abundance declines with advancing evolutionary stage, but the mean N abundance does not change much
along the RGB evolution. \citet{carbon82} found that the mean nitrogen abundance is essentially the same among subgiant
and asymptotic giant branch stars in M~92.
In the light of these results, a deeper investigation of N abundance changes along the RGB is needed.
This goes beyond the scope of this paper.

\subsection{$s$-process elements}\label{ELEMENTI_S}
Based on Figure~\ref{TEMPERATURA}, we note that Sr and Ba abundance values appear to be clustered around two 
distinct values for each RGB group. Confirming the results illustrated in Section~\ref{IND}, 
we found that R-RGB stars are enhanced in both Sr and Ba abundances with respect B-RGB stars. 
This appears immediately evident in Figure~\ref{SPETTRO}, where two representative B- and R-RGB stars 
are shown. These stars have essentially the same stellar parameters yet all these spectral features differ 
strongly\footnote{Accidentally Figure~\ref{SPETTRO} also confirms that the R-RGB star 05569 is a cluster member,
because its spectrum is perfectly identical to that of the B-RGB star 04837 apart from the molecular features and the $s$-process element 
absorption lines.}, the R-RGB stars clearly displaying strong NH, CN, and CH 
absorptions together with enhancement in the Sr and Ba abundances.

\begin{figure}
\includegraphics[width=\columnwidth]{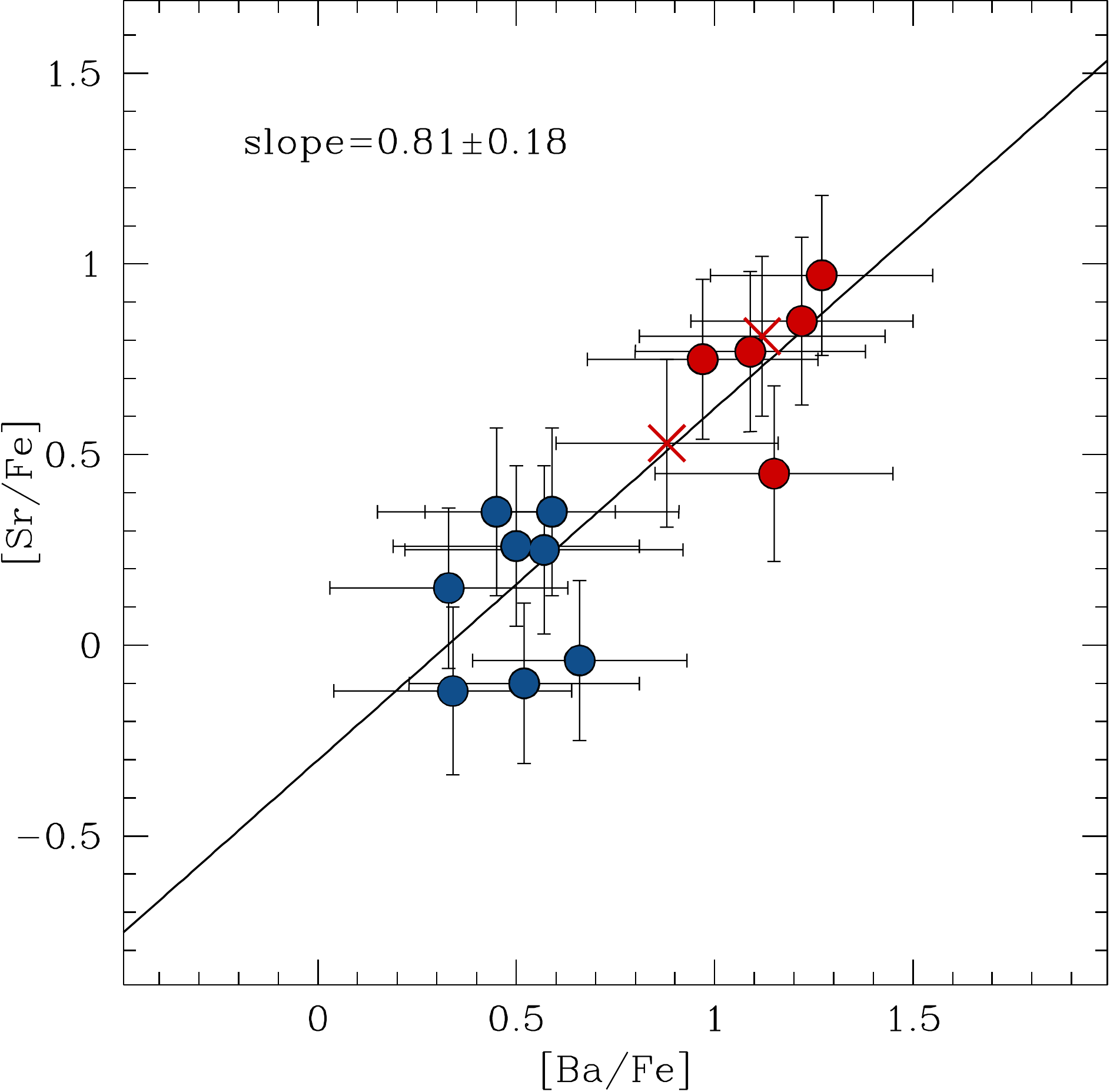}
\caption{[Ba/Fe] vs. [Sr/Fe] abundance ratios are plotted with typical measurement uncertainties. 
The black line is the best least squares fitting straight lines. The symbols are the same as in Figure~\ref{CN}.}
        \label{STRONZIO}
   \end{figure}

B-RGB stars have [Sr/Fe] $\simeq$ 0.25 $\pm$ 0.19 dex and [Ba/Fe] $\simeq$ 0.52 $\pm$ 0.07, that significantly 
differ from the average value found among the stars located on the R-RGB (for comparison [Sr/Fe] $\simeq$ 0.77 $\pm$ 0.16
and [Ba/Fe] $\simeq$ 1.12 $\pm$ 0.12 dex).
To better visualize this result, in Figure~\ref{STRONZIO}, 
we show [Ba/Fe] as a function of [Sr/Fe], showing a clear correlation between these two $s$-process elements. 

Figure~\ref{STRONZIO} also shows that we can isolate a $s$-process element rich group and a $s$-process element poor one, 
each clearly corresponding to R- and B-RGB stars, respectively.

The bimodality in $s$-process elements found in M~2 clearly resembles the cases of NGC~1851
\citep{yong08,villanova10,carretta1851,gratton1851} and M~22 \citep{marino09,dacosta11,marino11}. 
Beyond the classical C-N and Na-O anti-correlations, these GCs display bimodal heavy
element distribution and small metallicity spreads. Additionally, each $s$-group has its own
C-N, Na-O anti-correlation \citep{carretta1851,villanova10,lardo12,gratton1851,marino09,marino11,marino12},
suggesting that their star formation history must be more complicated than for the {\em normal} GCs.
Interestingly enough, both clusters display a very complex CMD, with splits at the level of the RGB
\citep{han09,carrettaSTR,gratton1851,marino11} and SGB \citep{milone08,piotto09,lardo12,piotto12,marino12}.
% Also for this cluster, is it natural to extend the the presence of the two groups of stars with different $s$-process element contents
% to the recently discovered split SGB \citep{piotto12}. 
% 
 \section{Summary and conclusions}\label{CONCLUSIONI}
We presented low resolution spectroscopy of 15 giants in M~2, located on the double
RGB in $V, U-V$ diagram, with the goal to chemically characterize these two groups of RGB stars. 
We derived reliable abundances for four key elements -- 
namely carbon, nitrogen, barium and strontium; thanks to the high quality of the MODS data.

As a first pass, we measured a set of spectral indices for all the observed stars.
B-RGB stars have S3839 (CN band) indices comparable with those measured in our previous work (L12), while R-RGB stars
have on average higher CN and CH absorptions. 
% We identify the presence of a CH-CN anti-correlation for both groups (Figure~\ref{CN}) albeit with a small
% statistical sample for the red RGB group.
The higher resolution (R@4000~\AA $\simeq$ 1000) of MODS spectra allowed us
also to provide reliable index measurements for the Sr II $\lambda$4077 and Ba II $\lambda$4554 lines. 
We found that the R-RGB group has, in general, higher indices than the B-RGB one (see Figure~\ref{CN}).

Second, we used spectral synthesis to measure C, N, Ba and Sr abundances for all our stars.
While B-RGB stars have carbon and nitrogen abundances well comparable to those derived from DOLORES data, 
R-RGB stars have enhanced [C/Fe] and [N/Fe] abundances (see Figure~\ref{ANTI}).
Also, R-RGB giants show a hint of C-N anti-correlation.
Moreover, we found that B- and R-RGB stars have different $s$-process element content 
(see Figure~\ref{STRONZIO}), R-RGB stars appearing systematically enhanced in both [Sr/Fe] and [Ba/Fe] 
than blue ones. 

Finally, we conclude that this $s$-process bimodality along the RGB could be associated -- as for M~22 
and NGC~1851 -- to the split SGB recently discovered by \citet{piotto12}. 
Moreover, the R-RGB could tentatively be linked to the group of stars at the faint, blue end of the HB: 
the relative frequency on the R-/B- RGBs roughly matches the relative frequency of 
these blue HB stars with respect to the whole HB population ($\sim$ 5\% of stars, see also L12).

The general picture emerging from our analysis confirms that M~2 shows a strong resemblance to M~22 and NGC~1851, as 
speculated in L12. 
This assumption is mostly based on (i) the observed $s$-process element bimodality among RGB stars with the 
$s$-rich group having on average higher C and N abundances, (ii) the presence of a double RGB and (iii) a split 
SGB. Interestingly enough, both NGC~1851 and M~22 show a small spread in their iron content \citep{yong08, carretta1851,gratton1851, 
marino12}. It would be of great interest to obtain high resolution spectra of to investigate  
possible iron spread among M~2 RGB population.

As a final comment, we want to stress that with this study we identified as an additional RGB a group of stars that 
could easily be mistaken as field stars.
Such situation could be more common than presently thought and a more careful inspection of $U$ based CMDs is 
highly advisable. One of the obvious next steps would be to obtain carefully constructed CMDs -- which include 
ultraviolet filters, for the GCs for which \citet{piotto12} reported the presence of a split SGB.
Such RGB additional sequences, if observed in other clusters, would present new challenges for  
multiple population studies.

\section{acknowledgements}
We acknowledge the support from the LBT-Italian Coordination Facility for the
execution of observations, data distribution and reduction.
 We thank the anonymous referee for a careful reading of the paper and helpful comments.
This research has made use of the SIMBAD database, operated at CDS, Strasbourg, France and of 
NASA Astrophysical Data System.
\bibliographystyle{mn2e}
%\input{MNRAS-replace}
%\bibliography{bibliography.bib}

\label{lastpage}
\end{document}